# Quantum spectroscopy of plasmonic nanostructures

Dmitry A. Kalashnikov, Zhenying Pan, Arseniy I. Kuznetsov, and Leonid A. Krivitsky*

Data Storage Institute, Agency for Science Technology and Research (A-STAR),

5 Engineering Drive I, 117608, Singapore

*Corresponding author: Leonid_Krivitskiy@dsi.a-star.edu.sg

We use frequency entangled photons, generated via spontaneous parametric down conversion, to measure the broadband spectral response of an array of gold nanoparticles exhibiting Fano-type plasmon resonance. Refractive index sensing of a liquid is performed by measuring the shift of the array resonance. This method is robust in excessively noisy conditions compared with conventional broadband transmission spectroscopy. Detection of a refractive index change is demonstrated with a noise level 70 times higher than the signal, which is shown to be inaccessible with the conventional transmission spectroscopy. Use of low photon fluxes makes this method suitable for measurements of photosensitive bio-samples and chemical substances.

**Introduction**

New insights in understanding the behavior of materials at the nano-scale and progress in nanofabrication capabilities have triggered great interest in the study of plasmonic nanostructures. One practical application of these structures is the development of sensing technology. Plasmonic sensors benefit from their bio-compatibility and high sensitivity [1-4]. With the sensitivity of the sensors approaching the atto-Molar scale, as well as for sensing of photosensitive substances, it is important that the analyte is not damaged or modified by the probing light. This brings the requirement of using extremely small doses of light, possibly down to the level of single photons, and highly sensitive equipment for the readout of the sensor response. For sensing with low photon counts, the main practical issue is careful suppression of the influence of the background noise. This is challenging with conventional spectroscopy but quantum optics can be exploited to effectively address this issue [5- 7].

Interfacing quantum optics with plasmonics is an area of active research. Plasmonic nanostructures are used as efficient optical antennas for increasing the collection efficiency of light, emitted by solid state single photon sources [8-16]. They are also shown to preserve quantum entanglement [17-19] and squeezing [20, 21], thus making it possible to use them in sub-wavelength quantum optical networks [22].

In this work, we demonstrate advantages of using frequency entangled photons, generated via spontaneous parametric down conversion (SPDC), in revealing the spectroscopic response of a plasmonic refractive index sensor under noisy conditions. The technique operates at a single photon level and is more robust to environmental noise, compared to conventional transmission spectroscopy.

**Quantum spectroscopy**

In the SPDC a pump photon travelling in a medium with quadratic susceptibility occasionally splits into a pair of photons, referred to as signal and idler [23, 24]. Signal and idler photons are created almost simultaneously (within tens of *fs*) with their energies and momenta obeying the conservation laws:

$$\omega_p = \omega_s + \omega_i \qquad \vec{k}_p = \vec{k}_s + \vec{k}_i, \qquad (1)$$

where $\omega_{p,s,i}$, and $\vec{k}_{p,s,i}$ are the energies and the wave vectors of the pump, signal and idler photons respectively. Signal and idler photons are detected with two single photon avalanche photodiodes (APD). Their pairwise correlation is measured by an electronic coincidence circuit, which produces a signal (coincidence) if APD counts arrive within the fixed time window typically of about several *ns*. Dedicated techniques, allow obtaining the widths of the SPDC spectra in a given spatial mode in the range of several hundreds of nanometers [25-28].

The sample under investigation is placed in the path of one of the photons (idler) and the frequency selection is performed over another photon (signal) by a monochromator [29-33], see Fig. 1(a). Since the sum of energies of the down-converted photons is equal to the energy of the pump, spectral selection over the signal photon defines the frequency of the correlated idler photon. Following [29], the number of coincidences between two APDs is proportional to the spectral function of the sample $T(\omega)$ at the frequency of the idler photon

$$R(\omega_s) \sim T(\omega_p - \omega_s). \qquad (2)$$

In contrast to the conventional transmission spectroscopy, the spectral selection is performed in the channel without the structure under test. Following the analogy with imaging experiments, we refer to the method as quantum (ghost) spectroscopy.

The remarkable feature of the quantum spectroscopy is its robustness against the environmental noise, which allows conducting measurements at extremely low photon fluxes. The noise typically includes background optical noise and electronic noise of the APDs. Let us denote the number of SPDC photocounts as $S_{s,i} = \eta_{s,i} P$, where $P$ is the number of photon pairs, produced by the SPDC source, and $\eta_{s,i}$ is the quantum efficiency of signal and idler channel, respectively. The number of coincidences is proportional to the probability of joint detection by the APDs in two channels:

$$R(\omega_s) = \eta_s \eta_i P. \tag{3}$$

Assuming that noise photocounts of the two APDs are uncorrelated, the number of noise coincidences $R_N$ is given by accidental overlap of photocounts within the time window of the coincidence circuit $\Delta t$. It also includes components due to accidental overlap of SPDC and noise photocounts:

$$R_N = N_s N_i \Delta t + S_s N_i \Delta t + N_s S_i \Delta t, \tag{4}$$

where $N_s$ and $N_i$ are the number of noise photocounts of APDs in signal and idler channels, respectively. The total number of coincidences is given by:

$$R_{Total} = R(\omega_s) + R_N. \tag{5}$$

Even though $N_s, N_i \gg S_s, S_i$, the number of noise coincidences $R_N$ is strongly suppressed compared to $R(\omega_s)$, because of a narrow coincidence time window $\Delta t$. Thus the spectral response of the structure can be revealed from $R_{Total}$ even under excessive noise situations.

**Comparison to the conventional transmission spectroscopy**

In conventional transmission spectroscopy, the sample is placed between the light source and the monochromator; the transmission spectrum is obtained directly from APD photocounts, see Fig. 1(b). The signal-to-noise ratio (SNR) for the transmission spectroscopy is given by $SNR_T = \eta_s P / N_s$. For the quantum spectroscopy SNR is given by $SNR_Q = R(\omega_s)/R_N = \eta_i \eta_s P / N_s N_i \Delta t$, (3, 4), where it is assumed that $N_s N_i \gg S_s N_i; N_s S_i$. The ratio of the two SNRs is given by:

$$SNR_Q / SNR_T = \eta_i / N_i \Delta t. \tag{6}$$

The advantage in the SNR for the quantum spectroscopy is provided by the use of high quantum efficiency, low noise APD in the idler channel, and the narrow coincidence time window $\Delta t$.

**The plasmonic array sensor**

We use an array of metal nanoparticles which has a narrow Fano-type plasmon resonance, due to diffractive coupling of localized surface plasmons [34-39]. The array acts as a metallic grating, and when the wavelength of incident light gets close to the Rayleigh cut-off wavelength, the diffracted wave propagates along the surface [40]. This effect is referred to as Wood-Rayleigh's anomaly. If the anomaly wavelength is close to the plasmon resonance of individual nanoparticles, then the collective plasmonic mode is excited. Excitation of the mode results in a narrow Fano-type resonance in the transmission spectrum of the array. Position and shape of the Fano resonance depend on the shape of the nanoparticles, distance between them, and refractive index of the surrounding medium. High sensitivity to the local change of the refractive index as well as high quality (Q) factor of the nanoparticle array resonance constitute the basis for its sensing applications [37, 41].

The nanoparticle array is fabricated using a combined method of nanosphere lithography with femtosecond laser-induced transfer which allows production of large-scale periodic arrays of spherical nanoparticles [37]. A hexagonal array with dimensions of approximately 1x1 mm$^2$ is fabricated with a particle diameter of 130 nm and a hexagonal lattice period of 1.1 µm, see Fig.2 (a, b). The fabricated particles are partially embedded (around two thirds) into a polydimethylsiloxane (PDMS) substrate having a part (around one third) above the substrate surface (this is accessible to the local environment). With the sample covered by second PDMS layer, the nanoparticles are in a homogeneous surrounding and the array provides Fano resonance centered at 806 nm with the quality factor Q ≈ 27, where Q is defined as the ratio of the resonance wavelength to the resonance width, defined from the fit using the Fano formula (see details below).

The refractive index sensitivity is measured by removing the cover PDMS layer and adding testing glycerin-water solutions with different concentration on top of the array. The sensitivity is calculated as a ratio of the resonance shift in *nm* to the change of the refractive index of the testing solution.

**Experimental setup**

The quantum spectroscopy setup is shown in Fig. 1(a). Three bulk BBO crystals (Dayoptics) with the thickness of 0.3 mm, 0.5 mm and 0.5 mm are pumped by a continuous wave vertically polarized diode laser at 407 nm (Omicron PhoxX-405-60). The BBOs are cut for type-I collinear SPDC. One of the BBOs is set for the frequency degenerate regime while another two are detuned from the degenerate regime by tilting their optical axis [26]. The resulting SPDC light source has a spectral width of ~250 nm, and is shown in Fig. 3. The pump is eliminated by a UV-mirror, and the SPDC is split by a non-polarizing beamsplitter (NPBS). In one arm of the NPBS we inserted a home-built diffraction grating monochromator (Thorlabs GR-1205) with resolution 1.5 nm. The light after the monochromator is detected by an avalanche photodiode (APD, Perkin Elmer SPCM-AQRH-14FC).

The sample is attached to a glass substrate and mounted in another arm of the NPBS in the focal plane of two confocal achromatic lenses with focus length $f$=60 mm. A mechanical translation stage (not shown) is used to position the sample. The focused light has a diameter of 300 $\mu m$ on the sample surface. The sample area with the array is surrounded by PDMS walls and covered by a thin glass plate forming a cell. The refractive index of the medium surrounding the array can be changed by adding different solvents inside the cell. The SPDC light passes through the broadband red glass filter (RG), and then it is coupled into the multimode fiber and detected by the APD (Perkin Elmer SPCM-AQRH-14FC). The signals from the two APDs are sent to the coincidence circuit with the time window of $\Delta t$=5 $ns$ (Ortec 567 TAC/SCA). The number of coincidences is measured as a function of wavelength, selected by the monochromator. The step size of the wavelength selection is 3 nm.

Noisy conditions are emulated by artificially adding photocounts of an additional APD (not shown), illuminated by a lamp, to the signal of the APD in the monochromator. This approach allows excluding the spectral dependence of the noise. The level of noise is controlled by changing the driving current of the lamp.

Results of the quantum spectroscopy are compared with those obtained by the transmission spectroscopy, see Fig. 1(b) for the experimental set-up. The light from the lamp after the single mode fiber (SMF) is sent through the sample under investigation, using the same optical system as described above, and then coupled into the monochromator. APD photocounts are recorded versus the wavelength, selected by the monochromator.

The experimental procedure in both cases includes transmission measurements of the samples substrate and the array. The resulting transmission spectrum is calculated as the ratio of the array and the substrate spectra.

**Results and discussion**

To ensure that the SPDC bandwidth fully covers the range of the structure's Fano resonance, we measure the sample transmission with the quantum spectroscopy set-up. The nanoparticle array is covered by a layer of PDMS to create homogeneous surroundings. The refractive index of the PDMS is about 1.4 [42]. The obtained spectrum is shown in Fig. 4. It reveals the Fano-type resonance with the distinctive asymmetrical shape. The gray solid curve in Fig.4 shows the fitting of the experimental results with the Fano formula [38, 39] with the following parameters: resonance wavelength $\lambda_R$ = 806 nm, resonance width $\Delta\lambda$ = 30 nm, and Fano (asymmetry) parameter F = -16. This gives the resonance quality factor Q = $\lambda_R/\Delta\lambda \approx 27$.

Refractive index sensing is demonstrated by depositing a few drops (~10 μl) of the testing solution of glycerin in water on the sample surface. In two different measurements the glycerin concentrations are 40% and 50% with corresponding refractive indexes of the solutions of 1.384 and 1.398, respectively (the refractive index values are taken from [43] for 20ºC). The refractive index of solution of 50% glycerin in water is close for refractive index of PDMS. Between subsequent measurements, the sample surface is washed with methanol and dried to remove the previous testing solution. Transmission spectra obtained at different glycerin concentrations using the conventional transmission spectroscopy are shown in Fig. 5(a). They demonstrate the red shift of the Fano resonance minima of about 8 nm when glycerin concentration changes from 40% to 50%. This corresponds to the sensitivity value of 570 nm/RIU, which is a characteristic value for refractive index sensing with plasmonic nanoparticles. Measurements performed by the quantum spectroscopy show similar results, see Fig. 5(b).

It is important to emphasize, that for the comparison of the two methods, all the measurements are performed with the same average flux of photons in the channel with the monochromator (in the vicinity of the resonance) equal to of $10^3$ photocounts/sec. The measurements are also performed for the same acquisition time: each point of the spectrum is obtained from the average of 20 measurements, each acquired for 20 s.

Results of the measurements performed under noisy conditions for the case of the conventional transmission spectroscopy are shown in Figs. 5(a, c, e). The noise levels are equal to $10^2$ photocounts/s in Fig. 5(a), $2*10^4$ photocounts/s in Fig. 5(c), and $7*10^4$ photocounts/s in Fig. 5(e). With an increasing noise level the resonance profile gets shallower and less pronounced. The maximum amount of noise at which it is feasible to perform sensing of the refractive index change is given by the interplay of the spectral function of the sample, the sensitivity of the sensor, the average photon flux, the acquisition time, and the efficiency of the detection system [5]. From the experimental data in Fig. 5(e) one can see that when the noise is about 70 times larger than the signal, see Fig. 5(e), the resonance positions for both glycerin concentrations completely overlap, making it impossible to observe any refractive index change.

The results obtained via the quantum spectroscopy, are shown in Figs. 5(b, d, f). The noise levels are equal to $10^3$ photocounts/s in Fig. 5(b), $2*10^4$ photocounts/s in Fig. 5(d), and $7*10^4$ photocounts/s in Fig. 5(f). In contrast to the transmission spectroscopy, the shape and the depth of the resonance do not significantly change. The resonance position remains distinguishable, even when the noise is 70 times larger than the signal in the monochromator, see Fig. 5(f). The only effect of the noise is a slight decrease in the accuracy of the resonance measurements due to contribution of noise coincidences. However, difference between the resonance minima remains the same as in the case of low noise. This shows that quantum spectroscopy has high resistance to noise and can be used with extremely low photon fluxes which are not possible with conventional transmission spectroscopy. This makes it very promising for sensing of photosensitive biological substances. Note that slight discrepancies in the shapes of Fano resonances in Figs. 5(a, b) are most likely caused by experimental uncertainties in setting of the angle of incidence of the sensing light on the sample.

Let us estimate the improvement in the SNR by using the quantum spectroscopy compared to the transmission spectroscopy using Eq. (6). In our experiment the width of the coincidence time window is set to $\Delta t = 5\ ns$. It is limited by the jitter of the APD (~400 *ps*) and the finite time resolution of the coincidence circuit (~3 *ns*). The noise of the APD in the idler channel is $N_i = 10^5$ photocounts/s. It is contributed by an optical noise, and photocounts due to detection of broadband idler photons at wavelengths, which are not correlated with wavelength of the signal photon, selected by the monochromator. Finally, the quantum efficiency in the idler channel is approximately $\eta_i = 5\ \%$. It accounts for the quantum efficiency of the APD ($\eta_{APD} = 50\ \%$), and optical losses in the idler channel. From

(6) we estimate the resulting improvement in the SNR using quantum spectroscopy $SNR_Q/SNR_T \approx 100$.

For experiments with actual light sensitive structures it would be advantageous to place the sample just after the monochromator in the signal beam. In this case the sample will be illuminated by a spectrally filtered light with fewer photons, compared to the present configuration. However, since the total number of detected photon pairs will remain the same there will be no improvement in the resulting SNR.

The ratio $SNR_Q/SNR_T$ can be significantly increased by using an additional monochromator in the idler arm. The two monochromators should select exactly conjugated frequencies of signal and idler photons, given by Eq. (1). It would be feasible to decrease $N_i$ by at least 2 orders of magnitude. Furthermore by using commercially available low-jitter APDs (~50 *ps*), and time-to-digital converters with timing resolution on the order of ~10 *ps* [44] it would be possible to set the coincidence time window to about ~100 *ps*. Given this, it is feasible to reach the ratio $SNR_Q/SNR_T$ as high as $5 * 10^5$.

**Conclusion**

In conclusion, a new method for refractive index sensing at extremely low photon fluxes and under excessive noisy conditions is demonstrated. The method combines the use of entangled photon pairs, generated via spontaneous parametric down conversion, and plasmonic nanoparticle arrays with Fano resonance. Pairwise correlation between the photons allows spectral measurements with cancellation of uncorrelated background optical noise, and electronic noise of the detectors.

Using this method, detection of a refractive index change of 0.014 is demonstrated with a noise level 70 times higher than the signal. The measured resonance shift of 8 nm corresponds to the relatively high refractive index sensitivity of ~570 nm/RIU provided by the Fano resonance of the nanoparticle array. It is shown that conventional spectroscopy fails to provide any sensing results in such excessive noisy conditions.

The developed approach will contribute to further progress in application of plasmonic sensors. In particular, it would allow non-disturbing measurements of ultra-small concentrations of chemicals (at atto-molar scale), and sensing of photo-sensitive substances, which can be affected by the probing light.

**Acknowledgments**

Manuel R. Gonçalves and Othmar Marti from the Institute of Experimental Physics, Ulm University are acknowledged for preparation of nanoparticle arrays on glass substrates by the nanosphere lithography. We acknowledge Carl Zeiss Pte Ltd (Singapore) for assistance with the SEM imaging of the sample. Fruitful discussions with Boris Luk'yanchuk, and Reuben Bakker are strongly appreciated.

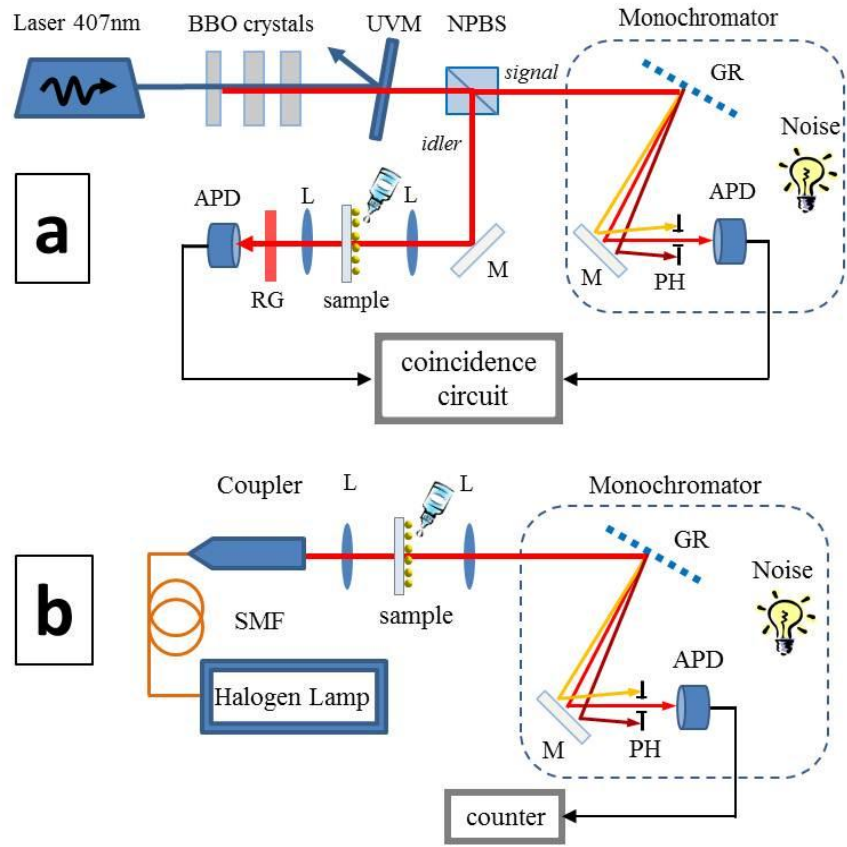

Fig. 1 (Color) (a) Quantum spectroscopy setup. Diode laser at 407 nm is used as a pump for a set of 3 BBO crystals cut for type-I SPDC; UV-mirror (UVM) reflects the pump beam and transmits the SPDC signal; A non-polarizing beam splitter (NPBS) splits the SPDC signal in two arms; In one arm there is a monochromator, which consists of a mirror (M), a diffraction grating (GR), a pin-hole, and an avalanche photodiode (APD). In another arm there is a sample, which represents an array of gold nanoparticles assembled into a cell. The SPDC signal is focused onto the sample by achromatic lenses (L), filtered by a red glass filter (RG), and detected by the APD. Counts of the two APDs are addressed to a coincidence circuit; source of noise is provided by the external APD, illuminated by the lamp (not shown), whose signal is mixed electronically with the signal of the APD in the signal channel. (b) Transmission spectroscopy setup. Halogen lamp is used as a source of the broadband light; the light after a single mode fiber (SMF) is collimated by a coupler, focused by lenses onto the sample, filtered by the monochromator and detected by the APD.

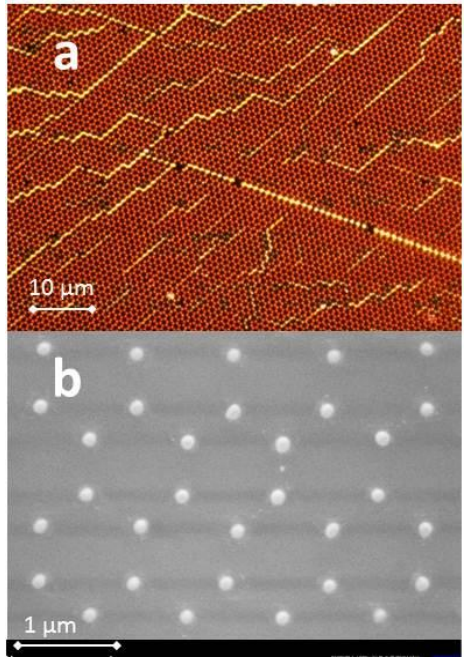

Fig 2. (Color) Images of the plasmonic nanoparticle array on PDMS substrate. (a) Dark-field microscope image at 10× magnification. (b) Scanning electron microscope (SEM) image.

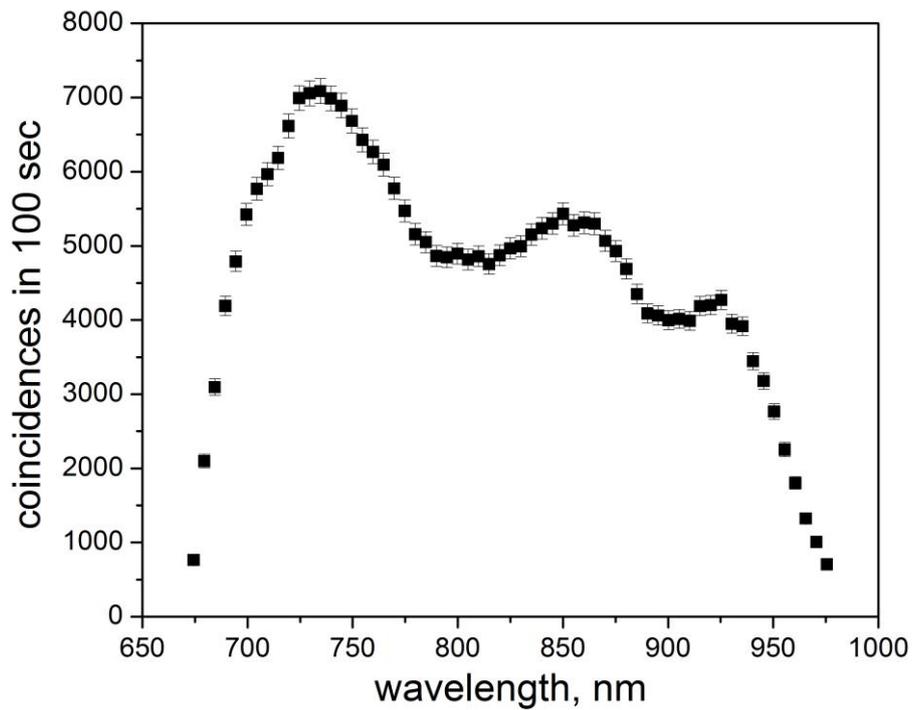

Fig. 3 Spectrum of coincidences of SPDC from 3 BBO crystals. The error bars are the standard deviations.

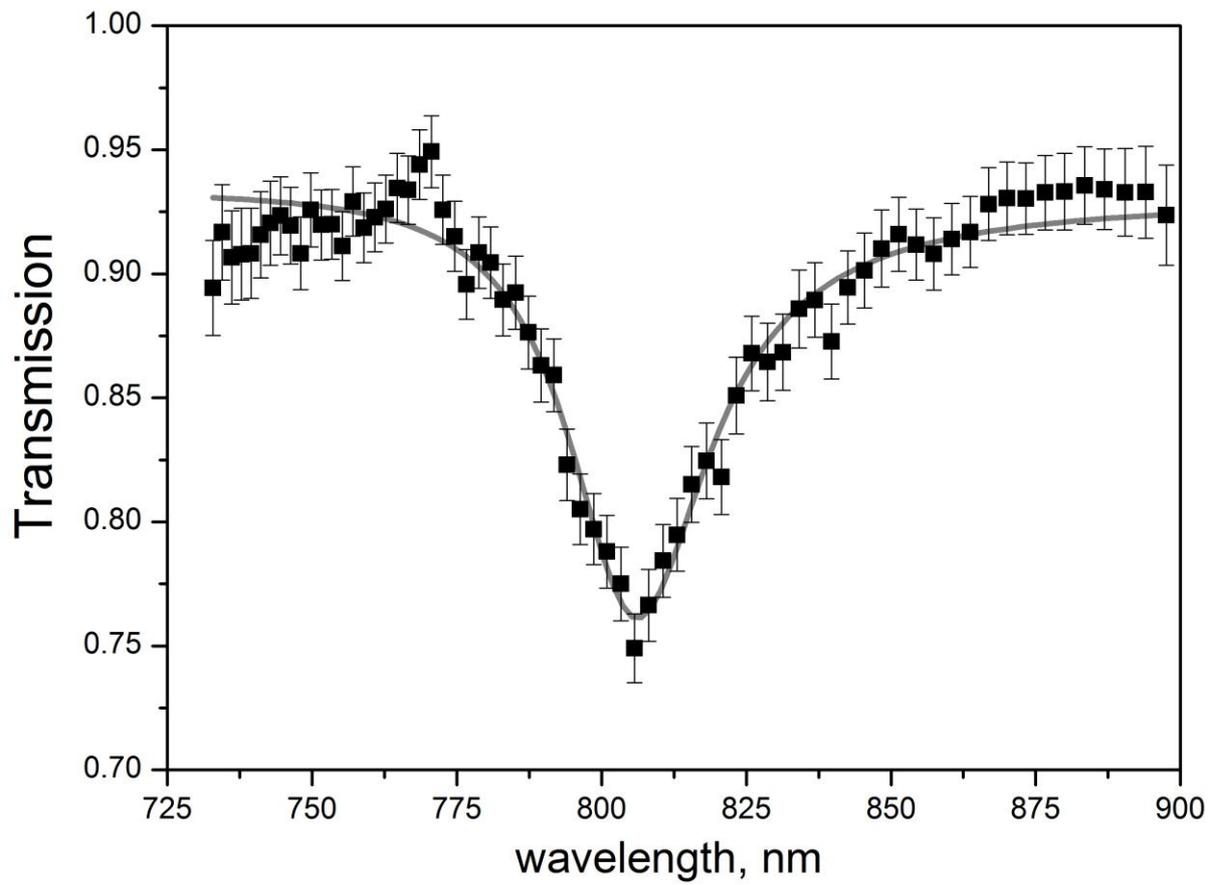

Fig. 4 Transmission spectrum of the nanoparticle array in homogeneous PDMS surroundings measured via quantum spectroscopy. Gray solid line is a fit with the Fano formula ($R^2$=0.94). The error bars are the standard deviations.

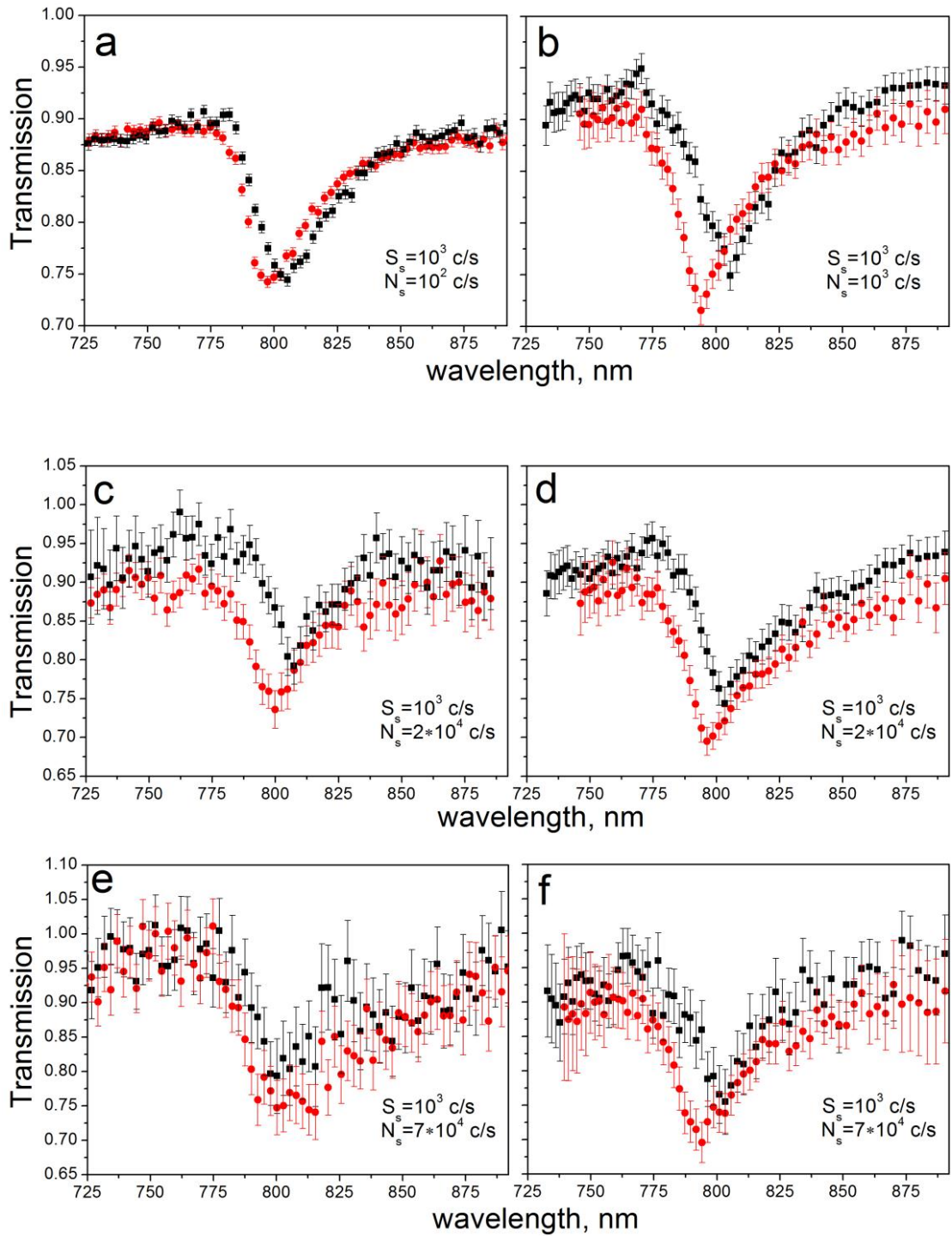

Fig. 5 (Color) Transmission spectrum of the nanoparticle array with different concentrations of glycerin-water solution on top, measured at different noise conditions. Red circles correspond to 40% of glycerin concentration and black squares are assigned to 50% of glycerin concentration. Results in (a), (c), and (e) are obtained by the conventional transmission spectroscopy with different noise levels of $10^2$ photocounts/s, $2*10^4$ photocounts/s, and $7*10^4$ photocounts/s, respectively. Results in (b), (d), and (f) are obtained by the quantum spectroscopy with different noise levels: $10^3$ photocounts/s, $2*10^4$

photocounts/s, and $7*10^4$ photocounts/s, respectively. The signal in the channel with the monochromator is the same for all the experiments in (a-f) and is equal to $10^3$ photocounts/s. The error bars are the standard deviations.